\documentclass[article]{ajs}
\usepackage{booktabs}          
\usepackage{tabularx}          
\usepackage{float}             
\usepackage{amsmath}           
\usepackage{mathtools}          
\usepackage{blkarray, bigstrut} 

\author{Guilherme P. Silva\\ Paran\'{a} Federal University \And
        Henrique A. Laureano\\ Pel\'{e} Pequeno Pr\'{i}ncipe Research Institute \AND
        Ricardo R. Petterle\\ Paran\'{a} Federal University \And
        Paulo J. R. J\'{u}nior\\ Paran\'{a} Federal University \And
        Wagner H. Bonat \\ Paran\'{a} Federal University
        }
\title{Multivariate Generalized Linear Mixed Models for Count Data}

\Plainauthor{Guilherme P. Silva, Henrique A. Laureano, Ricardo R. Petterle, Paulo J. R. J\'{u}nior, Wagner H. Bonat} 
\Plaintitle{A Capitalized Title: Something about a Topic} 
\Shorttitle{MGLMM count} 

\Abstract{
Univariate regression models have rich literature for counting data. However, this is not the case for multivariate count data. Therefore, we present the Multivariate Generalized Linear Mixed Models framework that deals with a multivariate set of responses, measuring the correlation between them through random effects that follows a multivariate normal distribution. This model is based on a GLMM with a random intercept and the estimation process remains the same as a standard GLMM with random effects integrated out via Laplace approximation. We efficiently implemented this model through the \texttt{TMB} package available in \texttt{R}. We used Poisson, negative binomial (NB), and COM-Poisson distributions. To assess the estimator properties, we conducted a simulation study considering four different sample sizes and three different correlation values for each distribution. We achieved unbiased and consistent estimators for Poisson and NB distributions; for COM-Poisson estimators were consistent, but biased, especially for dispersion, variance, and correlation parameter estimators. These models were applied to two datasets. 
The first concerns a sample from 30 different sites collected in Australia where the number of times each one of the 41 different ant species was registered; which results in an impressive 820 variance-covariance and 41 dispersion parameters being estimated simultaneously, let alone the regression parameters. The second is from the Australia Health Survey with 5 response variables and 5190 respondents. These datasets can be considered overdispersed by the generalized dispersion index. The COM-Poisson model overcame the other two competitors considering three goodness-of-fit indexes, AIC, BIC, and maximized log-likelihood values. As a result, it estimated parameters with smaller standard errors and a greater number of significant correlation coefficients. Therefore, the proposed model is capable of dealing with multivariate count data, either under- equi- or overdispersed responses, and measuring any kind of correlation between them taking into account the effects of the covariates.
}
\Plainkeywords{Laplace Approximation; Multivariate models; Optimization; Overdispersed count data; Regression models; Template Model Builder} 

\Pages{1--xx}

\Address{
  Guilherme P. Silva\\
  Laboratory of Statistics and Geoinformation\\
  Paran\'{a} Federal University\\
  CEP-81530-015 Curitiba (PR), Brazil\\
  E-mail: \email{guilhermeparreira.silva@gmail.com}\\
  URL: \url{https://github.com/guilhermeparreira}
}

\begin{document}


\section{Introduction}

It is well known that the Poisson distribution is the most popular model to deal with count data under the framework of the generalized linear models (GLM) \citep{glm}. However, it is limited to equidispersed count data, i.e., when the mean of the response variable is equal to the variance. As alternatives to the Poisson model, the statistical literature for univariate count data is rich, either from overdispersed or underdispersed perspectives. For example, negative binomial (NB) type II distribution parameterized on mean and dispersion parameter known as the quadratic parametrization of the variance \citep{winkelmannbook}, hurdle and zero-inflated models \citep{hurdlezi} and mixed Poisson regression \citep{winter2021poisson} are frequent options to model overdispersed count data; Gamma Count \citep{gammacount}, Conway-Maxwell-Poisson (COM-Poisson) \citep{compoisson} and the Extended Poisson Tweedie \citep{extendedPT} based on the Poisson Tweedie distribution \citep{pt1,pt2} can handle all three situations (under-, equi-, over-dispersed), even though the computation of their probability mass function (pmf) relies on numerical methods.

On the other hand, the bibliography for multivariate data is scarce. However, there is an increasing demand from researchers to analyze datasets with over one response variable. The benefit of it is to better investigate the relationship between the response variables \citep{wbonat.article}; at the same time that numerical methods can take advantage of it once more data is available to estimate model parameters.

Some methodologies to deal with multivariate data have been proposed and introduced by \citet{winkelmannbook} with a focus on building multivariate distributions. However, it comes with the price of some practical limitations. The multivariate Poisson-lognormal regression (MPLR) and the latent Poisson-normal regression models admit only overdispersed data. The multivariate Poisson-gamma mixture model (MPGM) and multivariate NB model (MNBM) are suitable only for overdispersed data with positive correlations. Another example is the copulas framework, which allows the building of multivariate distributions. But a negative correlation between many response variables is difficult to model, especially for count data \citep{copula}. \citet{multipoisson} proposed three alternatives to deal only with equi and overdispersed data constructed either via copulas or a mixture of distributions.

Distributions with no practical limitations were also developed. \citet{multgenpoisregmodel} proposed a multivariate generalized Poisson regression model that can deal with any kind of dispersion and correlation with an estimation based on the maximum likelihood (ML) paradigm. In a similar way, \citet{multicondpoisson} proposed a multivariate conditional Poisson regression model, where the relationship between response variables is measured by a coefficient in the linear predictor. In its turn, the dependence between response variables is conditional on the other response variables.

A very flexible modeling framework based on estimating functions, the Multivariate Covariance Generalized Linear Models (MCGLM) can also fit such data \citep{wbonat.pkg}. It uses only second-order moments assumptions and they estimated the parameters based on quasi-likelihood \citep{quasilikelihood}. It can accommodate correlated data based on an approach similar to the generalized estimating equations (GEE) \citep{gee}, allowing both multivariate responses and correlated data. We can also cite methodologies based on Bayesian inference, such as MCMC Generalized Linear Mixed Models via \texttt{MCMCglmm} package \citep{mcmcglmm} and Bayesian Regression Models using Stan - \texttt{brms} package \citep{brms}, which comes with the price of a greater computational time. We are not going to discuss Bayesian models in this article. 

Another alternative is to model the correlation between response variables in the same individual using the class of hierarchical GLM \citep{hglm}. This class allows to model of correlated variables or individuals via a random effect, an unobserved variable, that can follow any distribution. When the distribution of the random effect is Gaussian, we have the Generalized Linear Mixed Models (GLMM). However, GLMM is widely known and used to model the correlation between sample units, not for response variables, such a method is implemented in consolidated packages in software \texttt{R} \citep{rsoft}, such as glmmTMB \citep{glmmTMB}, lme4 \citep{lme4} and nlme \citep{nlme}.

In this article, we propose to model multivariate count data under the framework of GLMMs to accommodate the correlation between response variables. Parameter estimates are obtained through the maximum likelihood method \citep{maximumlikelihood}. We use multivariate normally distributed random effects to accommodate the correlation between response variables. The estimation process is similar to the one for GLMM and was implemented in \texttt{R} \citep{rsoft} through \texttt{TMB} package \citep{tmb}.

Here, we will focus on multivariate overdispersed data, whereas \citet{me_underdispersed} presented a similar approach focused on underdispersed data. This modeling approach for multivariate count data is evaluated for three different distributions of the response variables: Poisson, NB, and COM-Poisson mean parameterized \citep{compoissonmeanparam}. Even though COM-Poisson does not belong to the exponential family, we refer to it as a GLMM framework, once the estimation process remains the same regardless of the distribution being used.

This article contains six sections, including this introduction. Section 2 describes the datasets used to provide illustrative applications of the model. Section 3 proposes the multivariate generalized linear mixed model (MGLMM) model. Section 4 shows the result of the simulation study to assess the estimators' properties. Section 5 presents the results of the model applied to the datasets presented in Section 2. Finally, Section 6 discusses the major contributions of this article and future work.

\section{Data sets}
\label{cap:datasets}
%
%
\subsection{Dataset I: Australian Health Survey}
\label{cap:ahs}
%
%
The AHS is the largest survey conducted in Australia concerning health issues. The data used here is available through the \texttt{mcglm} package \citep{mcglm_pkg_2018} in the \texttt{ahs} object. The main aim of this study is to investigate whether more access to health care services and demographic covariates such as sex, age, and income are related to the number of times patients use health services.

The data has 5190 respondents and 15 variables; of which 10 were considered covariates and 5 as count response variables of a cross-sectional study with individuals over 18 years. The five response variables are Ndoc (Number of consultations with a doctor or specialist), Nndoc (Number of consultations with health professionals), Nadm (Number of admissions to a hospital, psychiatric hospital, nursing or convalescence home in the past 12 months), Nhosp (Number of nights in a hospital during the most recent admission) and Nmed (Total number of prescribed and non-prescribed medications used in the past two days). Table S1 in the supplementary material gives the description of each covariate.

\begin{figure}[H]
\vspace{0.35cm}
\setlength{\abovecaptionskip}{.0001pt}
\includegraphics[width=0.95\textwidth]{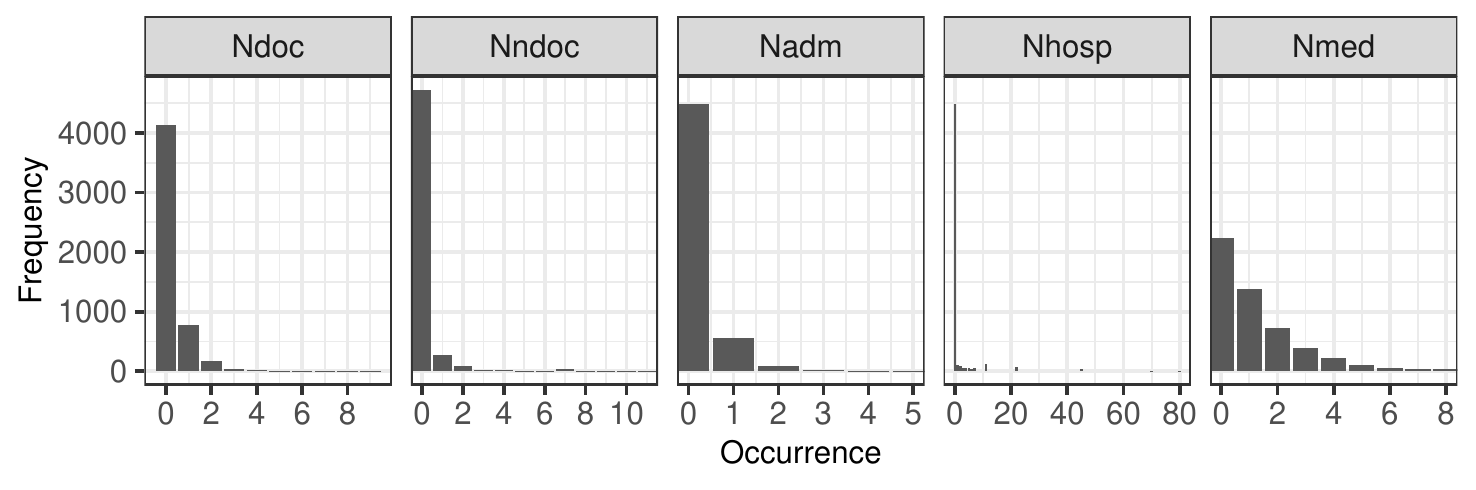}
\centering
\caption{Barplot for the count of each response variable from Australian Health Survey (AHS) data. Ndoc (Number of consultations with a doctor or specialist), Nndoc (Number of consultations with health professionals), Nadm (Number of admissions to a hospital, psychiatric hospital, nursing or convalescence home in the past 12 months), Nhosp (Number of nights in a hospital during the most recent admission) and Nmed (Total number of prescribed and non-prescribed medications used in the past two days).}
\label{fig:descahs}
\end{figure}

\autoref{fig:descahs} shows the barplot for each response variable. The figure suggests that all variables have some overdispersion, because of the right heavy-tailed and perhaps some excess of zeros. We highlight that such interpretations are marginal in the sense it does not take into account the effects of the covariates.
\autoref{tab:descAHS} presents some descriptive statistics for the 5 response variables. It is seen a positive (small in most cases) correlation between the 5 response variables, which is acceptable once they were measured in the same individual; except between Nadm and Nhosp, where the correlation is 0.996. Moreover, the mean and variance relationship between the variables shows overdispersion for all variables, which Nhosp being the most overdispersed. This is characterized by Fisher Dispersion Index (DI) greater than 1 \citep{dispersionindex}. Also, the generalized dispersion index (GDI) \citep{generalizedDI} classifies this dataset as overdispersed once it is greater than 1 and its 95\% confidence interval does not contain zero. However, these results should be confirmed by model fitting.

\begin{table}[H]
\caption{\label{tab:descAHS}Spearmean correlation, mean, variance, and dispersion index (DI) for the Australian Health Survey (AHS) response variables. The generalized dispersion index (GDI) and standard error (SE) equal 17.944 (1.99). Ndoc (Number of consultations with a doctor or specialist), Nndoc (Number of consultations with health professionals), Nadm (Number of admissions to a hospital, psychiatric hospital, nursing or convalescence home in the past 12 months), Nhosp (Number of nights in a hospital during the most recent admission) and Nmed (Total number of prescribed and non-prescribed medications used in the past two days).}
\centering
\begin{tabular}[t]{lccccccc}
\toprule
\multicolumn{1}{c}{ } & \multicolumn{4}{c}{Spearman Correlation $\rho$} & \multicolumn{1}{c}{Mean} & \multicolumn{1}{c}{Variance} & \multicolumn{1}{c}{DI} \\
\cmidrule(l{3pt}r{3pt}){2-5} \cmidrule(l{3pt}r{3pt}){6-6} \cmidrule(l{3pt}r{3pt}){7-7} \cmidrule(l{3pt}r{3pt}){8-8}
  & Nndoc & Nadm & Nhosp & Nmed & &  &  \\
 \cmidrule(l{3pt}r{3pt}){1-8}
Ndoc & 0.11 & 0.158 & 0.160 & 0.290 & 0.302 & 0.637 & 2.111 \\
Nndoc &  & 0.107 & 0.113 & 0.165 & 0.215 & 0.932 & 4.341 \\
Nadm &  &  & 0.996 & 0.135 & 0.174 & 0.258 & 1.484 \\
Nhosp &  &  &  & 0.139 & 1.334 & 37.455 & 28.083 \\
Nmed &  &  &  &  & 1.218 & 2.423 & 1.989 \\
\bottomrule
\end{tabular}
\centering
\end{table}

%
%
\subsection{Dataset II: Abundance ant species}
\label{cap:ant}
%
This data was obtained from a study conducted in south-eastern Australia by \citet{ant}, and it is available in the software R \citep{rsoft} throughout the \texttt{mvabund} package \citep{mvabund}. The study consisted of counting the number of 41 different ant species that fell into pitfall traps during 18-day sessions in 30 different sites in south-eastern Australia between November 2007 and April 2008. The primary interest of this dataset is to investigate the relationship between the environmental variables with the occurrence of different ant species and the co-occurrence of ant species.
%
The response variables considered are the occurrence of each of the 41 species at the 30 sites, while the covariates comprise 5 environmental variables and we gave their full description in the supplementary material Table S2. The name of the response variables starts with an index number (1,...,41) followed by the abbreviated name of each species.



\begin{figure}[H]
\vspace{0.35cm}
\setlength{\abovecaptionskip}{.0001pt}
\includegraphics[width=0.95\textwidth]{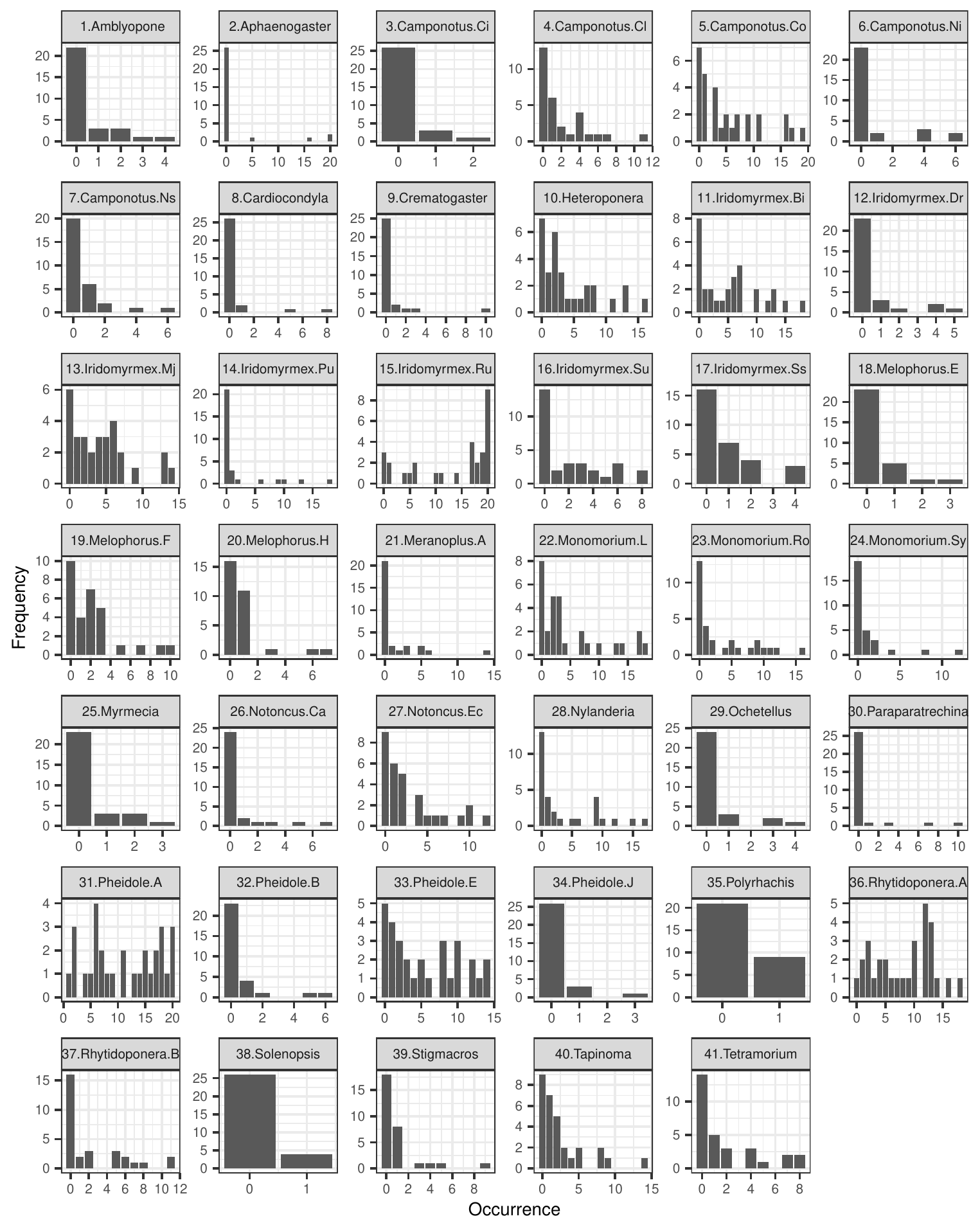}
\centering
\caption{Barplot for the occurrence of each ant species genus from ANT data.}
\label{fig:barant}
\end{figure}

\autoref{fig:barant} presents the barplot for each response variable from ANT data. The 41 different species present high variability. Some species were only seen one time in a single site, such as \textit{Polyrhachis} and \textit{Solenopsis}, while \textit{Iridomyrmex} and \textit{Pheidole} were seen 20 times on over one site.

\autoref{tab:tabant} presents the mean, variance, and DI for every response variable and the GDI for the dataset. The variance is higher than the mean almost for all variables, except for \textit{Polyrhachis} and \textit{Solenopsis}; which is described by a DI>1. This dataset is classified as overdispersed based on the GDI index = 11,543 and the lower bound of a 95\% confidence interval was greater than 1.

\begin{table}[H]

\caption{\label{tab:tabant}Mean, variance, dispersion index (DI) for each genus for ANT data. Generalized dispersion index (GDI) and its standard error (SE) equals 11.54 (.92)}
\centering
\begin{tabular}[t]{lrrrlrrr}
\toprule
Response & Mean & Variance & DI & Response & Mean & Variance & DI\\
\midrule
1.Amblyopone & 0.533 & 1.085 & 2.034 & 21.Meranoplus.A & 1.333 & 8.713 & 6.534\\
2.Aphaenogaster & 2.033 & 32.999 & 16.229 & 22.Monomorium.L & 4.733 & 32.409 & 6.847\\
3.Camponotus.Ci & 0.167 & 0.213 & 1.276 & 23.Monomorium.Ro & 3.433 & 20.944 & 6.100\\
4.Camponotus.Cl & 1.933 & 7.099 & 3.672 & 24.Monomorium.Sy & 1.167 & 6.902 & 5.916\\
5.Camponotus.Co & 5.300 & 33.252 & 6.274 & 25.Myrmecia & 0.400 & 0.662 & 1.655\\
\addlinespace
6.Camponotus.Ni & 0.867 & 3.430 & 3.958 & 26.Notoncus.Ca & 0.633 & 2.654 & 4.191\\
7.Camponotus.Ns & 0.667 & 1.816 & 2.724 & 27.Notoncus.Ec & 2.900 & 12.300 & 4.241\\
8.Cardiocondyla & 0.500 & 2.879 & 5.759 & 28.Nylanderia & 3.733 & 25.720 & 6.889\\
9.Crematogaster & 0.567 & 3.633 & 6.412 & 29.Ochetellus & 0.433 & 1.082 & 2.496\\
10.Heteroponera & 4.067 & 19.857 & 4.883 & 30.Paraparatrechina & 0.700 & 4.976 & 7.108\\
\addlinespace
11.Iridomyrmex.Bi & 5.333 & 26.437 & 4.957 & 31.Pheidole.A & 11.100 & 40.300 & 3.631\\
12.Iridomyrmex.Dr & 0.600 & 1.834 & 3.057 & 32.Pheidole.B & 0.567 & 2.047 & 3.613\\
13.Iridomyrmex.Mj & 4.300 & 15.803 & 3.675 & 33.Pheidole.E & 5.467 & 22.809 & 4.172\\
14.Iridomyrmex.Pu & 2.033 & 20.447 & 10.056 & 34.Pheidole.J & 0.200 & 0.372 & 1.862\\
15.Iridomyrmex.Ru & 13.300 & 59.045 & 4.439 & 35.Polyrhachis & 0.300 & 0.217 & 0.724\\
\addlinespace
16.Iridomyrmex.Su & 2.133 & 6.809 & 3.192 & 36.Rhytidoponera.A & 8.300 & 25.528 & 3.076\\
17.Iridomyrmex.Ss & 0.900 & 1.610 & 1.789 & 37.Rhytidoponera.B & 2.400 & 11.834 & 4.931\\
18.Melophorus.E & 0.333 & 0.506 & 1.517 & 38.Solenopsis & 0.133 & 0.120 & 0.897\\
19.Melophorus.F & 2.133 & 6.740 & 3.159 & 39.Stigmacros & 0.967 & 3.826 & 3.958\\
20.Melophorus.H & 0.900 & 2.783 & 3.092 & 40.Tapinoma & 2.533 & 11.154 & 4.403\\
\addlinespace
 & & & & 41.Tetramorium & 1.933 & 7.030 & 3.636\\
\bottomrule
\end{tabular}
\end{table}

\begin{figure}[H]
\vspace{0.35cm}
\setlength{\abovecaptionskip}{.0001pt}
\caption{Correlogram of ANT species occurrence using spearman correlation}
\includegraphics[width=0.95\textwidth]{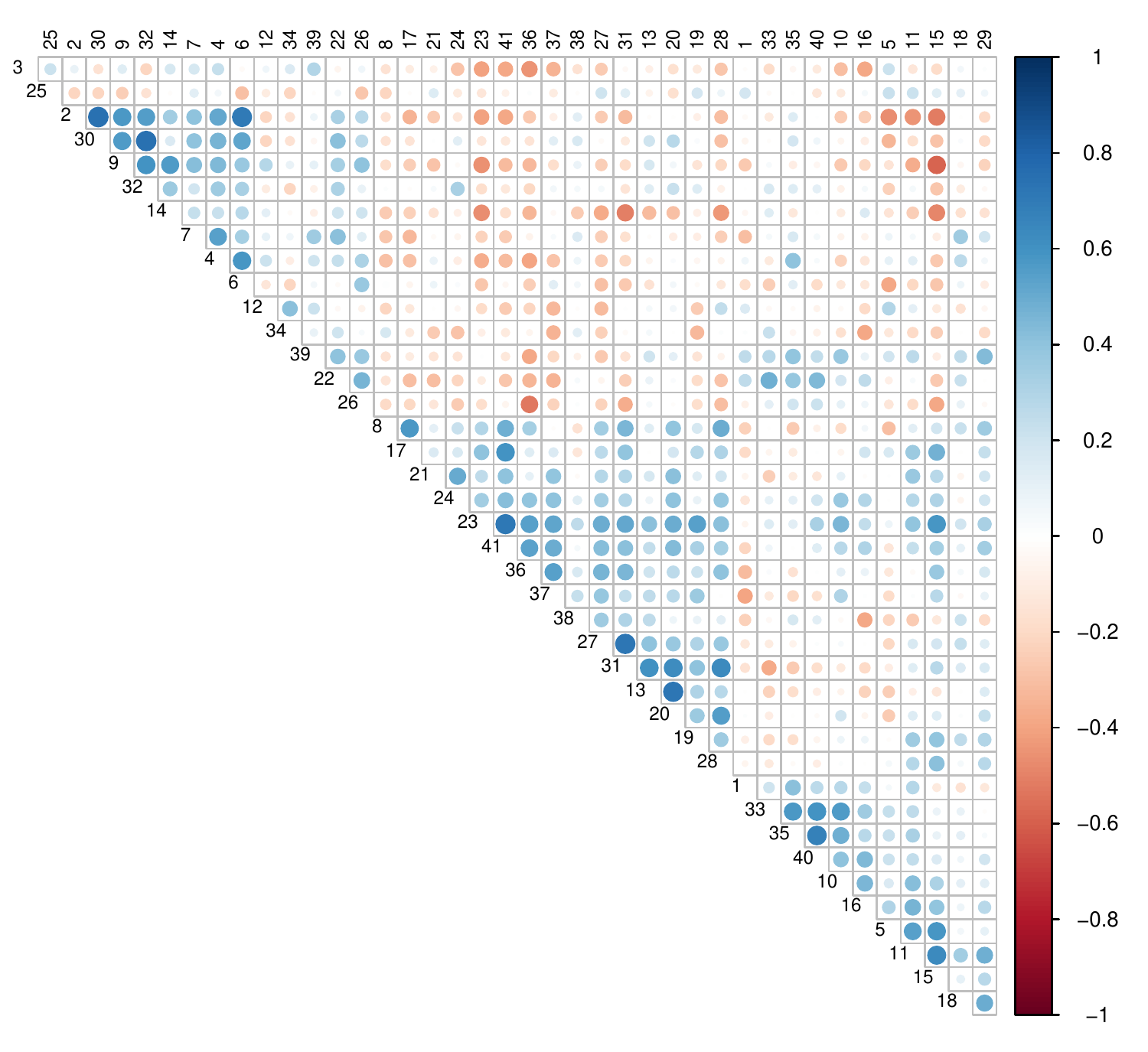}
\centering
\label{fig:corant}
\end{figure}

\autoref{fig:corant} explores how the occurrence of different species is related to each other by the Spearman correlation in a correlogram. The marginal correlation ranges from -.58 up to .74, which is well distributed along all potential values of the correlation parameter $\rho = \left ( -1,1 \right )$.

%
%



\section{Multivariate Generalized Linear Mixed Model (MGLMM) for Count Data}
\label{cap:modelo}
%
%

This section follows the same description in \citet{me_underdispersed} and presents the MGLMM model to completeness; even though the explanations are the same, but succinct. The fundamental difference is that in this article we will explore the model for overdispersed counts. In summary, we extended the standard GLMM \citep{glmm} for multiple responses.

Let $Y_{ij}$ be a random variable for $i=1,\dots,n$ independent subjects and $r=1,\dots,k$ response variables. Also, consider that $\mathrm{x}_{irj}$ for each subject \emph{i} and response variable \emph{r} is the value of the \emph{j-th} covariate of \emph{k} known covariates. We first specify a standard GLMM joint model with only a random intercept:

\begin{equation*}
\mathrm{Y}_{ir} \mid \mathrm{b}_{ir} \sim f(\mu_{ir};\phi_{ir}),
\end{equation*}
where the conditional distribution of $\mathrm{Y}_{ir}$ on the random effects follows a same distribution \emph{f} allowing different mean and dispersion. Next, the linear predictor:

\begin{equation*}
\label{eq:linear_predictor}
\mathrm{g}_r(\mu_{ir}) = x_{irj}^{\top}\boldsymbol{\beta}_{r} + \mathrm{b}_{ir},
\end{equation*}
where $\mathrm{g}_r(.)$ is a suitable link function, $\boldsymbol{\beta}_{r}$ is a $p \times 1$ vector of parameter estimates and $\mathrm{b}_{ir}$ is the random intercept value for subject \emph{i} and response \emph{r}. Finally, the random effects distribution is specified as:

\begin{equation*}
\begin{pmatrix}
\mathrm{b}_{i1}\\
\mathrm{b}_{i2}\\
\vdots \\
\mathrm{b}_{ir}
\end{pmatrix}
\sim \mathrm{NM} \begin{pmatrix}
\begin{bmatrix}
0\\
0\\
\vdots\\
0
\end{bmatrix};
\underset{r\times r}{\boldsymbol{\Sigma}} = \begin{bmatrix}
\sigma^{2}_{1} & \rho_{12}\sigma_1\sigma_2 & \dots & \rho_{1r}\sigma_1\sigma_r\\
\rho_{21}\sigma_2\sigma_1 & \sigma^{2}_{2}  & \dots & \rho_{2r}\sigma_2\sigma_r \\
\vdots & \vdots & \ddots  & \vdots \\
\rho_{r1}\sigma_r\sigma_1 & \rho_{r2}\sigma_r\sigma_2 & \dots & \sigma^{2}_{r}
\end{bmatrix}
\end{pmatrix}
,
\end{equation*}
where each random effect has mean 0, variance $\sigma^2$, and correlation $\rho_{r{r}'} (r\neq{r}')$ between each pair of random effects. This framework is general and can be applied to any distribution \emph{f} and link function $\mathrm{g}_r(.)$. Nevertheless, as we are dealing with count data, we considered only Poisson, binomial negative type II and COM-Poisson distributions, and a logarithm link function. Extending this model for different distributions for each response ($f_r$) is also possible, but we will not address it.

Maximum likelihood is the estimation procedure, and it is fully described in \citet{me_underdispersed}. We integrated out the random effects via numerical integration using Laplace approximation \citep{laplace}. The Newton-Raphson method was efficiently implemented to perform the inner optimization. We optimized the marginal likelihood using first-order derivatives methods, such as \texttt{BFGS} and \texttt{OPTIM} routines available in the \texttt{R} software. The derivatives of the joint and marginal likelihood were obtained through automatic differentiation \citep{ad}. This was efficiently implemented by the \texttt{TMB} package, which provides \texttt{C++} templates where an objective function must be supplied, in this case, the log-likelihood function. 

The variability of the random variable in this model is being measured by two parameters, the variance of the random effect and the dispersion of the pmf. A crucial point of the model is the great flexibility to learn these two types of variances with no identifiability problems.

The code used to produce the results of this paper is available on \url{https://github.com/guilhermeparreira/papers/tree/master/AJS_MGLMM_Overdispersed_Count} and \url{http://www.leg.ufpr.br/doku.php/publications:papercompanions}.

\section{Simulation Studies}

In this section, we present simulation studies to assess the properties of the MLE estimators (bias and consistency). We considered a bivariate regression model for count data. We designed 12 simulation scenarios with four different sample sizes, 100, 250, 500, and 1000, and three different correlations between random effects, $\rho = -0.5, 0, 0.5$. For the regression structure, we considered only an intercept for each response, with $\beta_{01} = \text{log}(7)$ and $\beta_{02} = \text{log}(1.5)$. The variance of random effects were $\sigma^{2}_{1} = .3$ and $\sigma^{2}_{2} = .15$. The dispersion parameter for NB and COM-Poisson was equal to $\phi = 1$ and $\nu = .7$ respectively, which induces a small overdispersion.

We generated 150, 200, and 300 datasets for Poisson, COM-Poisson, and NB distribution for each design. The primary idea was to generate 100 datasets for each distribution. However, as it did not return the SE of the parameter estimates in every repetition, it was necessary to increase the number of datasets generated proportionally to the number of SE failures for each distribution to obtain at least 100 valid estimations. The following three subsection presents the results for Poisson, NB, and COM-Poisson scenarios.

\subsection{Poisson scenario}
\label{cap:poisson_sim}

\begin{figure}[H]
  \centering
  \includegraphics[width=\textwidth]{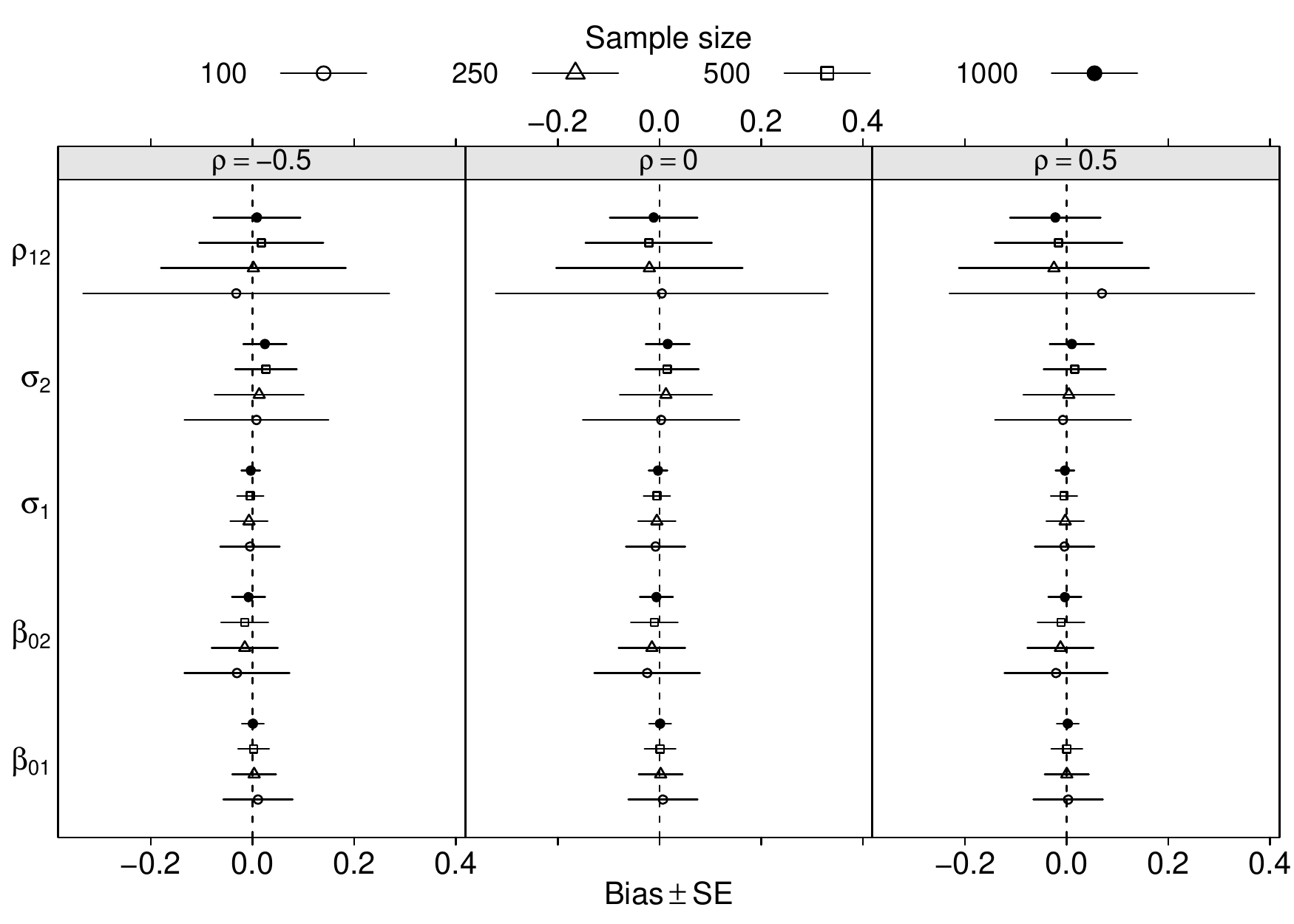}
  \caption{{Average bias and its 95\% confidence interval based on the average standard error (SE) by four different sample sizes, three correlation coefficients for the bivariate Poisson regression model. The true value for each parameter were with $\beta_{01} = \text{log}(7)$, $\beta_{02} = \text{log}(1.5)$, $\sigma^{2}_{1} = .3$ and $\sigma^{2}_{2} = .15$.}}\label{fig:poissonbias}
\end{figure}

\autoref{fig:poissonbias} presents the average bias and its confidence interval based on the average SE by sample size and simulation scenario for the bivariate Poisson regression model. It shows that, for all simulation scenarios, both the average bias and SE tend to be 0 as the sample size increases. This suggests the consistency and unbiasedness of the MLE estimator for large samples.

\begin{figure}
\vspace{0.05cm}
\centering
\setlength{\abovecaptionskip}{.0001pt}
\includegraphics[width=\textwidth]{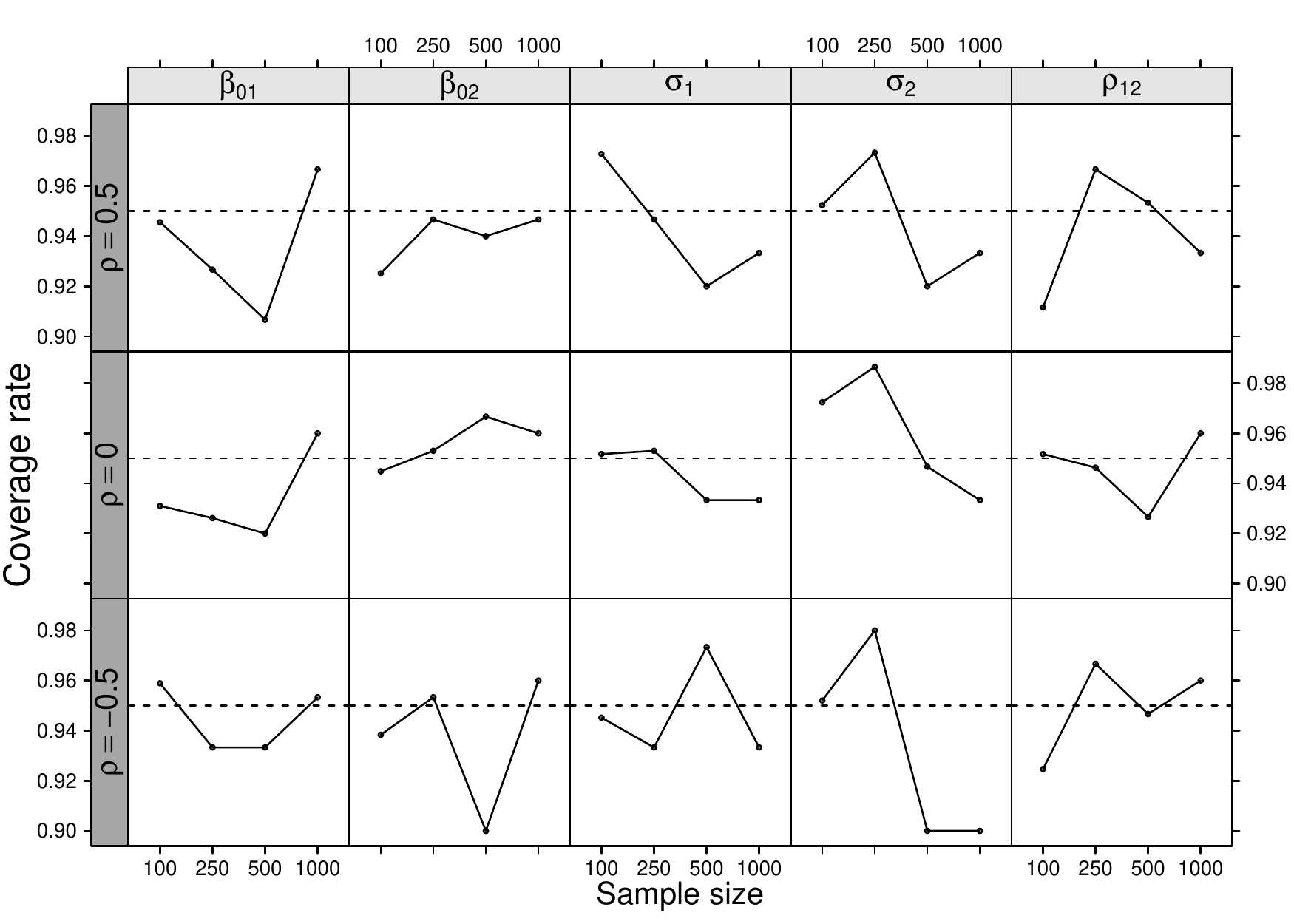}
\caption{{Coverage rate for each parameter (in the columns) by sample size and correlation coefficient for the Poisson regression model. The true value for each parameter were $\beta_{01} = \text{log}(7)$, $\beta_{02} = \text{log}(1.5)$, $\sigma^{2}_{1} = .3$ and $\sigma^{2}_{2} = .15$.}}\label{fig:poisson_cov}
\end{figure}

\autoref{fig:poisson_cov} presents the coverage rate for each parameter by sample size and simulation scenario for the bivariate Poisson regression model. Overall, all empirical coverage rates are close to the nominal level of 95\%, varying between 90\% and 98\% approximately. In special, the coverage rate of regression parameters $\boldsymbol{\beta}_{0r}$ are slightly greater than the nominal level. For the variance of random effects $\boldsymbol{\sigma}_{r}$, the coverage rate is slightly smaller than the nominal level. Finally, for the correlation between random effects $\rho$, the coverage rate is slightly lower when $\rho = .5$, and slightly greater when $\rho = \{0, -0.5\}$ compared to the 95\% nominal level.

Even for the Poisson distribution, which is the simplest case to estimate because there is no dispersion parameter, there were 13 out of 1800 (3$\times$4$\times$150) simulations that did not return SE for some parameters estimates or produced extreme values due to large SEs that were not considered into the results. We used the \texttt{PORT} algorithm to estimate the model because it was more stable and faster than \texttt{BFGS} in most situations. It also happened in \citet{tmb}, where 9 study cases from different model settings ranging from linear regression to multivariate stochastic volatility models were considered, and \texttt{PORT} had a better performance than \texttt{BFGS}.
\subsection{Negative Binomial scenario}
\label{cap:nb_sim}
\begin{figure}[H]
  \centering
  \includegraphics[width=\textwidth]{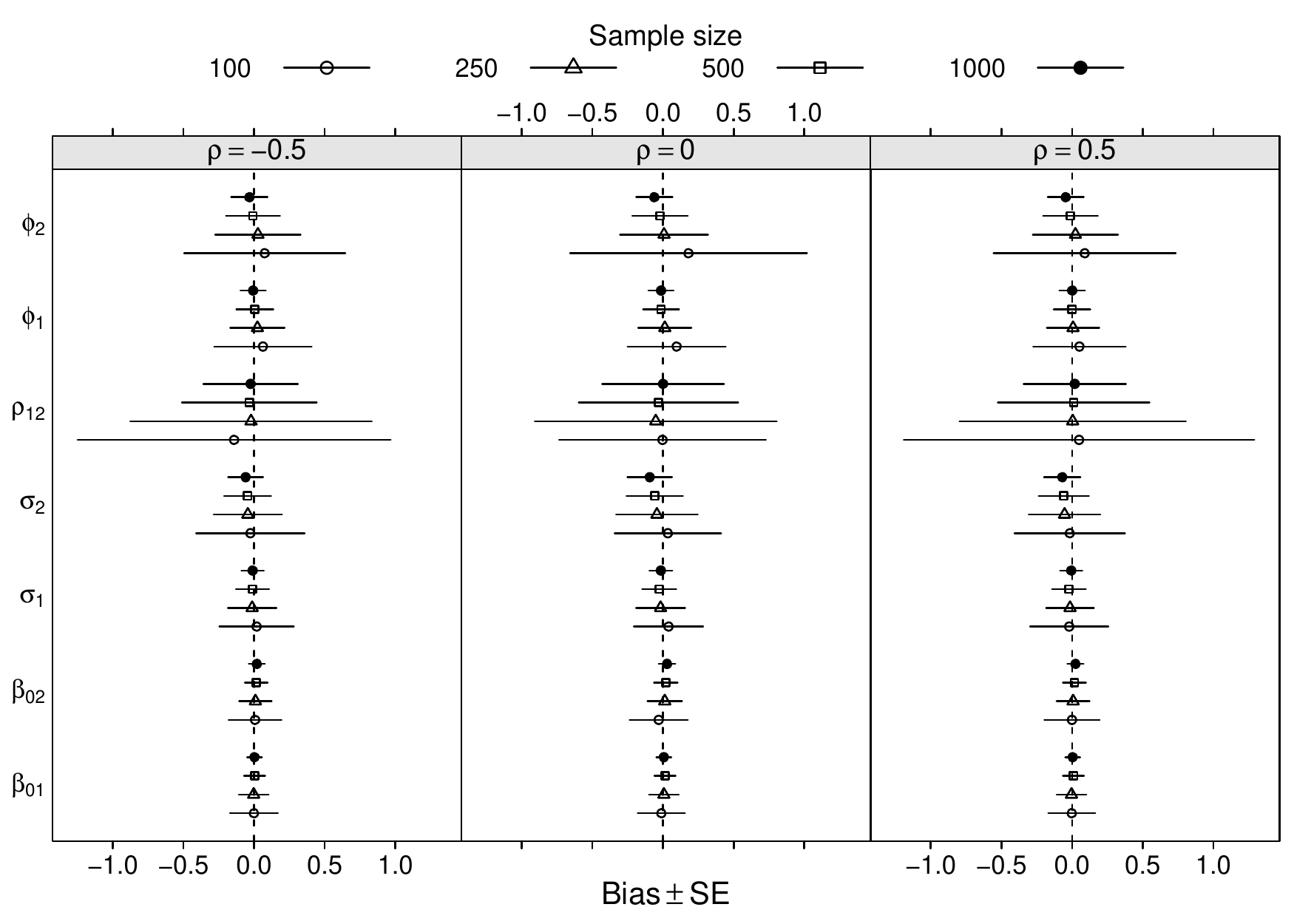}
  \caption{Average bias and its 95\% confidence interval based on the average standard error (SE) by four different sample sizes, three correlation coefficients for the bivariate negative binomial (NB) regression model. The true value for each parameter were $\beta_{01} = \text{log}(7)$, $\beta_{02} = \text{log}(1.5)$, $\phi = 1$, $\sigma^{2}_{1} = .3$ and $\sigma^{2}_{2} = .15$.}\label{fig:nb_bias}
\end{figure}

\autoref{fig:nb_bias} suggests that for all simulation scenarios, both the average bias and SE tend to be 0 as the sample size increases. This suggests the consistency and unbiasedness of the MLE estimator for large samples. Concerning the variance of the random effect $\sigma_1$ and $\sigma_2$ were slightly underestimated.

\begin{figure}[H]
  \centering
  \includegraphics[width=\textwidth]{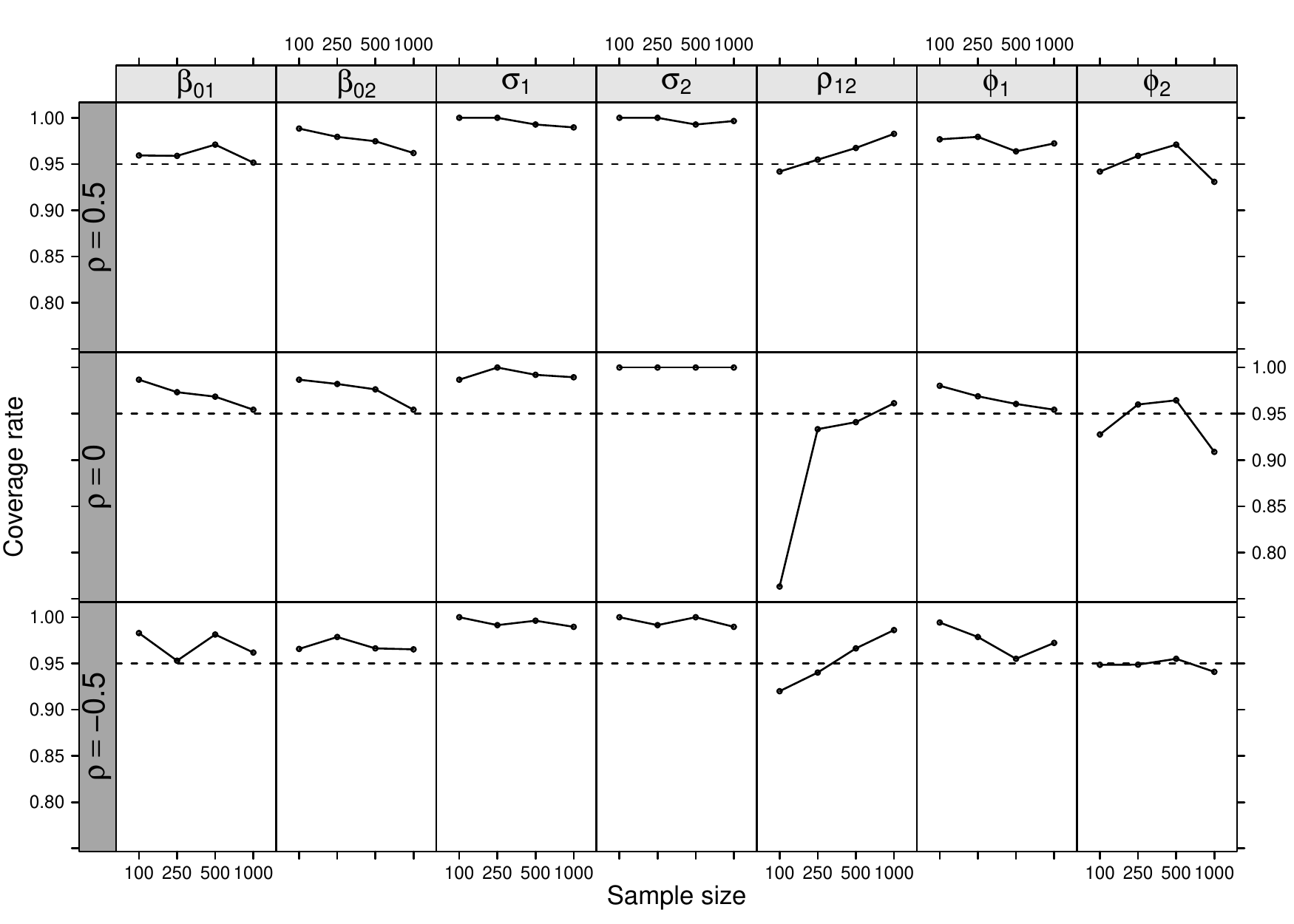}
  \caption{Coverage rate for each parameter (in the columns) by sample size and correlation coefficient for bivariate negative binomial (NB) regression model. The true value for each parameter were $\beta_{01} = \text{log}(7)$, $\beta_{02} = \text{log}(1.5)$, $\phi = 1$, $\sigma^{2}_{1} = .3$ and $\sigma^{2}_{2} = .15$.}\label{fig:nb_cov}
\end{figure}

Overall, \autoref{fig:nb_cov} shows that the empirical coverage rates are close to the nominal level of 95\%. In particular, the coverage rate of regression parameters $\boldsymbol{\beta}_{0r}$, the variance of random effects $\boldsymbol{\sigma}_{r}$ and the correlation between random effects $\rho$ are slightly greater than the nominal level. On the opposite, there was a coverage rate close to 80\% for $\rho$ estimated when the sample size was equal to 100 and $\rho = 0$.

Estimation problems were more severe for the bivariate NB regression model compared to the Poisson and occurred in 740 out of 3600 (3$\times$4$\times$300) simulations (20,6\%). It was also necessary to remove those iterations when the dispersion parameter $\phi$ was greater than 5 from the results (we simulated at 1).

\subsection{COM-Poisson scenario}
\label{cap:cmp_sim}

\begin{figure}[htb]
  \centering
  \includegraphics[width=\textwidth]{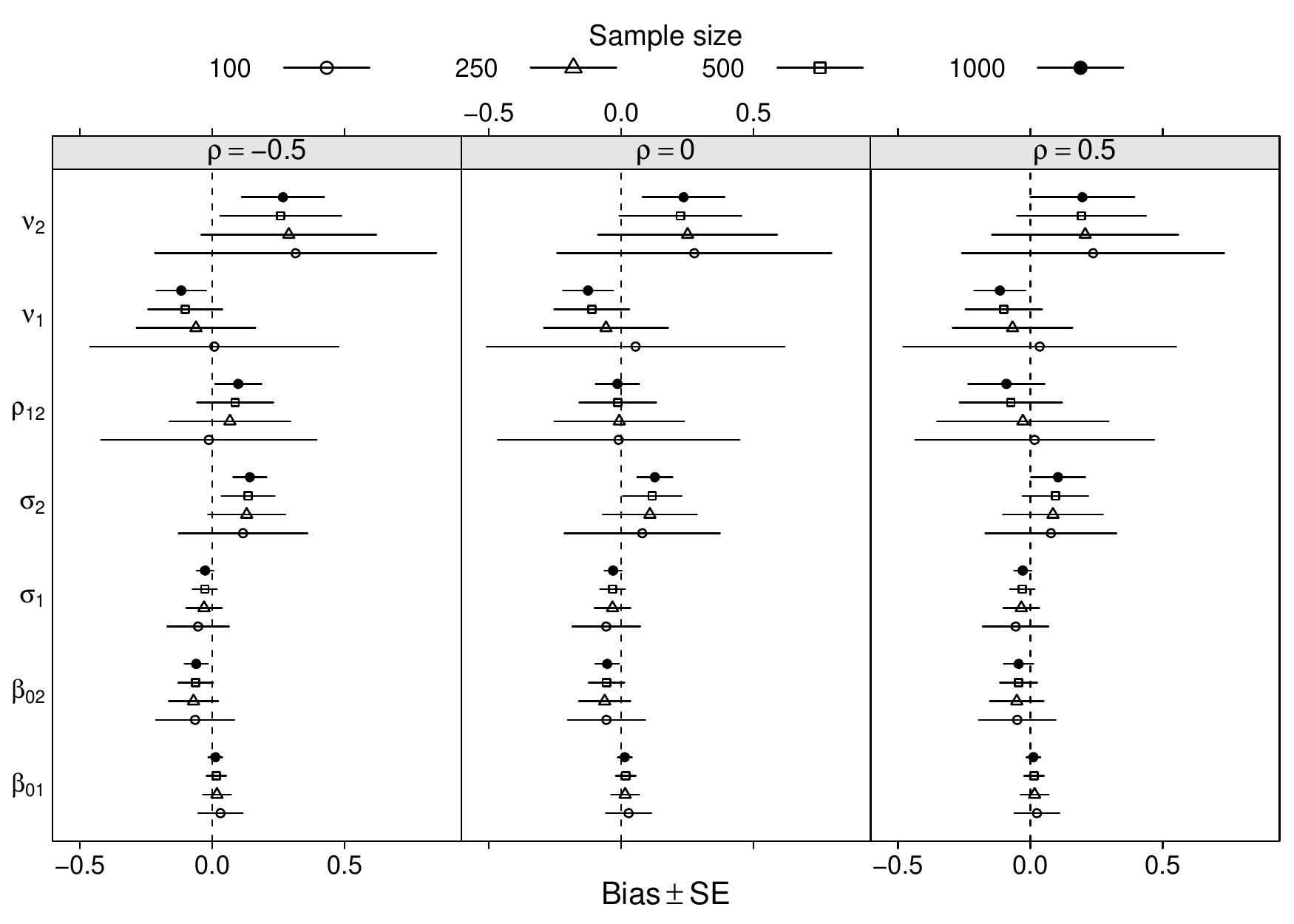}
  \caption{Average bias and its 95\% confidence interval based on the average standard error (SE) by four different sample sizes, three correlation coefficients for the bivariate COM-Poisson regression model. The true value for each parameter were $\beta_{01} = \text{log}(7)$, $\beta_{02} = \text{log}(1.5)$, $\nu = .7$, $\sigma^{2}_{1} = .3$ and $\sigma^{2}_{2} = .15$.}\label{fig:cmp_bias}
\end{figure}

\autoref{fig:cmp_bias} shows that, for all simulation scenarios, the SE tends to be 0 as the sample size increases, but the bias does not. This suggests the non-consistency of the MLE estimator. Moreover, almost every estimate is biased. While $\nu_2$ is overestimated, $\nu_1$ is underestimated (and the bias does not decrease with an increase in sample size). The correlation parameter $\rho$ has a typical behavior: when the data was simulated with $\rho=0$ there was almost no bias, for $\rho=.5$ a negative bias and for $\rho=-.5$ a positive bias; the model forces the correlation parameter towards zero. The standard deviation of random effect $\sigma_2$ is overestimated, while $\sigma_1$ is slightly underestimated. Regarding the regression parameters, $\beta_{02}$ is slightly underestimated and $\beta_{01}$ suggests a negligible bias. 

An interesting possible relationship among parameters is that when $\nu$ was overestimated (making the model more underdispersed: $\nu>1$), $\sigma$ was also overestimated (increasing the variance of the model). In contrast, when $\nu$ was underestimated (making the model even more overdispersed: $\nu<1$), $\sigma$ was also underestimated (decreasing the variance of the model). It seems that when $\nu$ makes the dispersion of the model greater, $\sigma$ makes the variance lower, and vice versa. It may be related because of a possible non-orthogonality between these parameters.

\begin{figure}[htb]
  \centering
  \includegraphics[width=\textwidth]{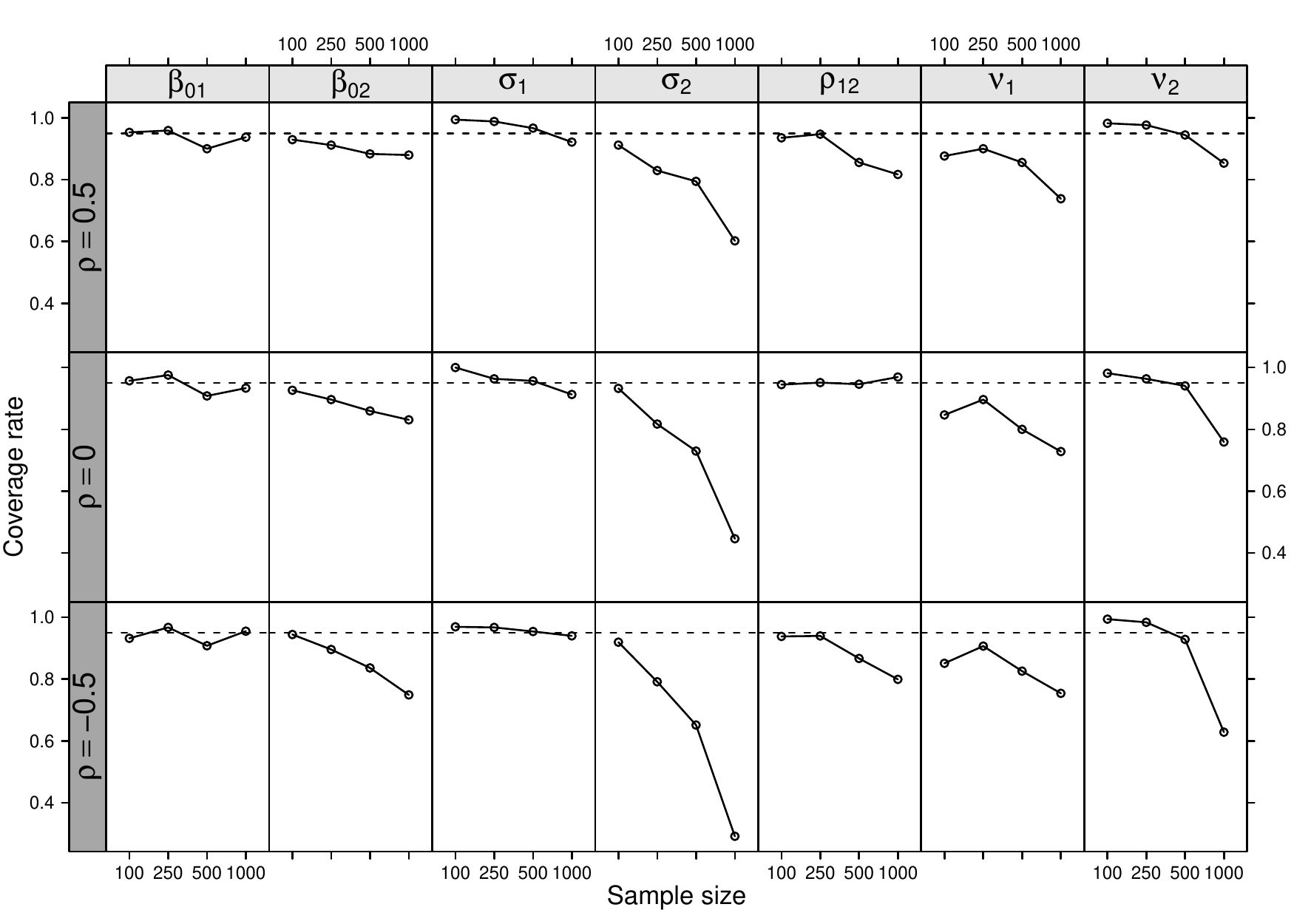}
  \caption{Coverage rate for each parameter (in the columns) by sample size and correlation coefficient for bivariate COM-Poisson regression model. The true value for each parameter were $\beta_{01} = \text{log}(7)$, $\beta_{02} = \text{log}(1.5)$, $\nu = .7$, $\sigma^{2}_{1} = .3$ and $\sigma^{2}_{2} = .15$.}\label{fig:cmp_cov}
\end{figure}

Overall, \autoref{fig:cmp_cov} shows that empirical coverage rates are not close to the nominal level of 95\%, which agrees with the results presented in \autoref{fig:cmp_bias}, where the bias did not decrease even for a higher sample size. In particular, the coverage rate of $\sigma_{2}$ had the worst results among all parameters (the coverage rate decreases as sample sizes increase), followed by $\nu_2$. Not surprisingly, these parameters had the two largest biases in \autoref{fig:cmp_bias}. In contrast, results for $\beta_{01}$, $\sigma_{1}$ and $\rho$ (when the data was generated with $\rho=0$) had coverage rates close to 95\% level.

Estimation problems were more severe for the COM-Poisson regression model compared to the Poisson and less severe if compared to NB, accounting for 245 (10,2\%) out of 2400 (3$\times$4$\times$200). Besides the rules used for Poisson to classify extreme values, it was necessary to remove those models when the dispersion parameter $\nu$ was greater than 4 and the SE of $\rho$ estimate was greater than 2.

\section{DATA ANALYSES}
\label{cap:dataanalyses}

This section presents the data analyses of the two datasets presented from \autoref{cap:datasets}. Initial values for the model in \autoref{cap:modelo} were chosen carefully in the following way. Firstly, and for every dataset, we fitted a Poisson model via quasi-likelihood in \texttt{MCGLM} package \citep{wbonat.pkg} to obtain initial parameter estimates for the regression and variance parameters (based on the variance of the residuals). Because of the difference in methodologies, we set the correlation parameter to 0. These were initial values for the Poisson model; the final estimates from the Poisson became the initial values for the NB model, and from the NB, the initial estimates of COM-Poisson. The initial values to the dispersion parameter were $\phi = 1$ for NB, i.e., small overdispersion, and $\nu = 1$, i.e., equidispersion to the COM-Poisson model.

The estimation process started with a simple random sample (SRS) of size 300 (seed 2390) for AHS data. The parameters obtained from the sample were used as initial values to fit the model to the whole dataset. For ANT data, we did not use this step because of the small size of the dataset (30). We then optimize the same model more than once iterating over optimizers. In the first round we used \texttt{PORT} routines and its final estimates were used as initial estimates to optimize the same model with \texttt{BFGS}; in its turn, the final estimates from \texttt{BFGS} were used as initial estimates to \texttt{PORT}; and another final round to \texttt{PORT}. We used this rationale because these models are complex to estimate and this procedure produced reasonable results.

Besides the full model presented in \autoref{cap:modelo}, we also tried to estimate four simpler versions to check whether the model was over-specified. We made the simplifications either by fixing values from the $\boldsymbol{\Sigma}$ or the dispersion parameter. The first one aimed to reproduce independent response variables setting the correlation parameter to 0. In the second version, we fixed $\phi = 1$ and $\nu=1.5$ for NB and COM-Poisson models inducing small overdispersion and underdispersion, respectively. In the third simplified model, we set the variance to $1$. The last strategy used only a single variance parameter to estimate all response variables. The last three simpler versions described aimed to answer whether we can estimate both dispersion and variance parameters in the same model. We only tested them for NB and COM-Poisson models since they have dispersion parameters.

The reported results are only from the full data and the best model version for each distribution after removing covariates that did not have the SEs estimated. We then presented the maximized log-likelihood value, Akaike and Bayesian Information Criterion (AIC and BIC), and the number of parameters estimated (np).

\subsection{AHS data}
\label{cap:resultAHS}

\autoref{tab:ahsfit} presents the goodness-of-fit measures for AHS data from different distributions and specifications.

\begin{table}[H]
\caption{\label{tab:ahsfit}Goodness-of-fit measures for AHS data from different distributions and specifications. Number of parameters estimated (np), Akaike and Bayesian Information Criterion (AIC and BIC), and log-likelihood value (Loglik).}
\centering
\begin{tabular}[t]{lccccc}
\toprule
Model & np & AIC & BIC & Loglik\\
\midrule
Poisson & 70 & 33814 & 34272 & -16837\\
NB & 75 & 33826 & 34318 & -16838\\
COM-Poisson & 75 & 33106 & 33598 & -16478\\
\bottomrule
\end{tabular}
\end{table}

The COM-Poisson full model was the best among all distributions and the simpler versions of each distribution.

\begin{figure}[H]
  \centering
  \includegraphics[width=\textwidth]{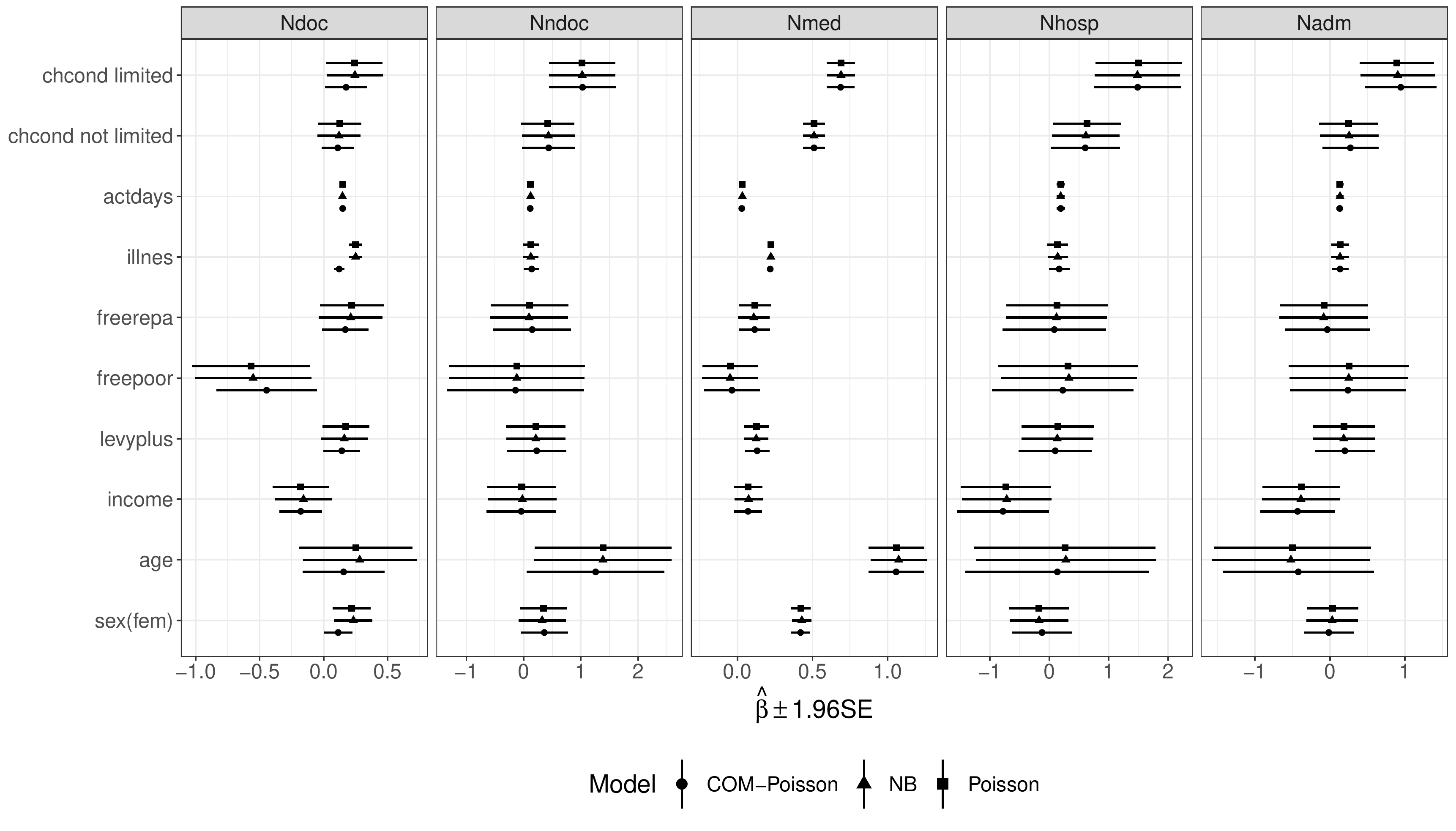}
  \caption{Regression parameter estimates and 95\% confidence intervals by outcome and final model for the Australian Health Survey (AHS) data.}\label{fig:ahsbeta1}
\end{figure}

\autoref{fig:ahsbeta1} presents the regression parameter estimates and 95\% confidence intervals by outcome and final model. We can see that the confidence interval amplitude is almost the same for all models, but slightly smaller for the COM-Poisson models over the competitors for almost all estimates for the Ndoc response variable; for the other response variables, we saw little differences. The point estimates were very close between Poisson and NB models, with a difference for the COM-Poisson model in some estimates, for example, freepoor, age, sex, illness, and chcond limited for Ndoc. We may relate this to the bias found in \autoref{fig:cmp_bias}.

\autoref{tab:ahsdisp} presents the dispersion estimates for each model and outcome. Even though this data can be considered as marginally overdispersed according to the $\text{GDI} = 17.94$ presented in \autoref{tab:descAHS}, we see that $\phi$ approaches the infinity (suggest an equidispersed model by NB distribution), and $\nu$ is greater than 1 for all response variables and indicates underdispersion. The expected result would be $\phi$ and $\nu$ approaching zero, which results in an overdispersion of both distributions. However, as explored in \autoref{cap:resultANT}, this behavior may occur because of the variance of the random effect. For the COM-Poisson model, the only $\nu$ estimate which indicates overdispersion was for Nmed, which had a sample DI equal to 1.99 (small overdispersion).

\begin{table}[H]
\caption{\label{tab:ahsdisp}{Dispersion of parameter estimates and standard errors (SEs) for each model and outcome of the Australian Health Survey (AHS) data}}
\centering
\begin{tabular}[t]{ccccc}
\toprule
\multicolumn{1}{c}{ } & \multicolumn{2}{c}{NB($\phi$)} & \multicolumn{2}{c}{COM-Poisson($\nu$)} \\
\cmidrule(l{3pt}r{3pt}){2-3} \cmidrule(l{3pt}r{3pt}){4-5}
Outcome & Estimate & SE & Estimate & SE\\
\midrule
Ndoc & 1.7e+04 & 1.8e+05 & 9.160 & 0.432\\
Nndoc & 5.9e+03 & 6.0e+04 & 2.837 & 0.336\\
Nmed & 3.6e+03 & 2.4e+04 & 0.674 & 0.032\\
Nhosp & 8.5e+03 & 1.2e+05 & 5.110 & 0.615\\
Nadm & 8.5e+03 & 8.1e+04 & 6.665 & 0.431\\
\bottomrule
\end{tabular}
\end{table}

We presented the correlation coefficient estimates and their SEs in parentheses in (1) for the COM-Poisson model. Among the 10 correlation coefficients returned, COM-Poisson had 6 significant correlation coefficients. The stars in the matrix represent significant coefficients at the 5\% level. Even though we do not have a primary interest in interpreting each correlation coefficient estimate, it is important to know which of them is significant, because it may be related to a smaller SE.

We can see an almost perfect significant correlation between Nadm and Nhosp, and a strong correlation between Nmed and Ndoc. We had also seen this pattern in \autoref{tab:descAHS}. 
\begin{multline}
$\footnotesize$
\begin{aligned}
\label{eq:corrmatrixahs}
  \text{COM-Poisson}_\rho&=
  \begin{blockarray}{*{5}{c} l}
    \begin{block}{*{5}{>{$\footnotesize}c<{$}} l}
      Ndoc & Nndoc & Nmed & Nhosp & Nadm & \\
    \end{block}
    \begin{block}{[*{5}{c}]>{$\footnotesize}l<{$}}
    1 & -0.07(0.03)^{*} & 0.78(0.09)^{*} & -0.02(0.02) & -0.02(0.02) \bigstrut[t]& Ndoc\\
     & 1 & 0.34(0.14)^{*} & -0.04(0.03) & -0.05(0.03) & Nndoc\\
     &  & 1 & 0.25(0.09)^{*} & 0.25(0.09)^{*} & Nmed\\
     &  &  & 1 & 0.99(<.01)^{*} & Nhosp\\
     &  &  &  & 1 & Nadm \\
    \end{block}
  \end{blockarray}
\end{aligned}
\end{multline}

\subsection{ANT data}
\label{cap:resultANT}

We presented the best model for each distribution in \autoref{tab:antfit2}. Here, we included two models for the COM-Poisson distribution because they were equivalent in the intermediate models based on the adopted criterion.

\begin{table}[H]

\caption{\label{tab:antfit2}Goodness-of-fit measures for ANT data from the best parametrization for each distribution. Number of parameters estimated (np), Akaike and Bayesian Information Criterion (AIC and BIC), and log-likelihood value (Loglik).}
\centering
\begin{tabular}[t]{lccccc}
\toprule
Model & np & AIC & BIC & logLik\\
\midrule
Poisson & 1057 & 5085.08 & 6566.15 & -1485.54\\
NB & 1098 & 5166.98 & 6705.50 & -1485.49\\
COM-Poisson & 1098 & 4965.87 & 6504.38 & -1384.93\\
Fixed dispersion COM-Poisson & 1057 & 5016.70 & 6497.77 & -1451.35\\
\bottomrule
\end{tabular}
\end{table}

From \autoref{tab:antfit2} we can see that the best models were from COM-Poisson distribution; Poisson and NB models had similar \texttt{logLik} results. The COM-Poisson models had higher \texttt{logLik} and \texttt{BIC}, and smaller \texttt{AIC} than the fixed dispersion model. A logLik ratio test (LRT) between these two COM-Poisson models resulted in $p<.00001$ ($\chi_{41}=132.84$), which gives evidence in favor of the fully specified model. Therefore, we will present the estimates of the full models for each distribution.

\begin{figure}[H]
  \centering
  \includegraphics[width=\textwidth]{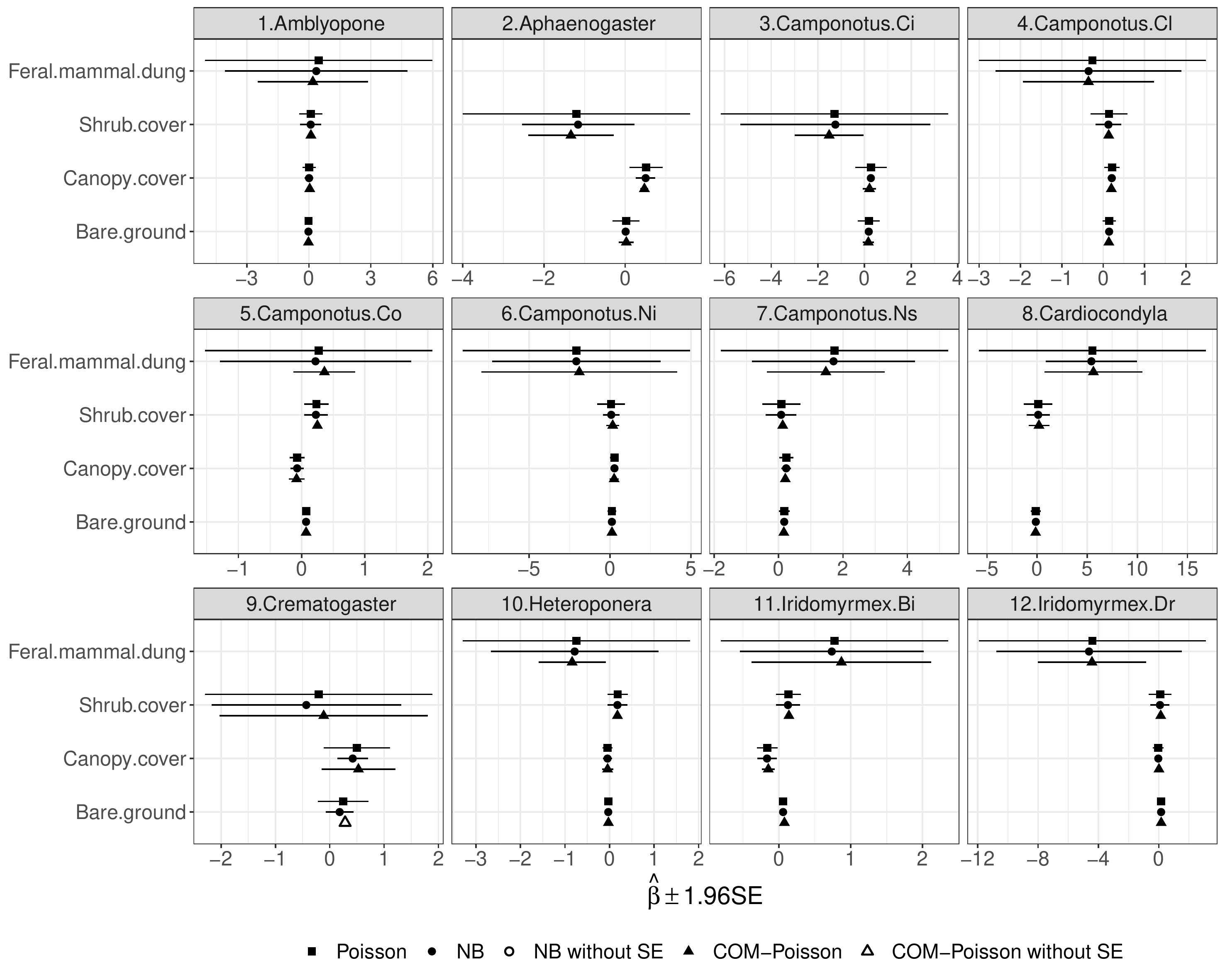}
  \caption{Regression parameter estimates and 95\% confidence intervals by outcome and final model for ANT data.}\label{fig:antbeta1}
\end{figure}

\autoref{fig:antbeta1} presents the regression parameter estimates and 95\% confidence intervals by outcome and final model for the first 12 response variables. We presented the same graphic for the remaining response variables in the supplementary material, Figures S1, S2, and S3, once we found similar patterns for all response variables. First, we can see that not all response variables share the same linear predictor (covariate feral mammal dung is not presented in the second, third and ninth response variables). Second, there were still some regression estimates' SEs that were not returned by the model and are presented by a hollow circle. It was necessary to make the distinction when the SE was returned and when it was tiny (filled circle, such as bare ground for COM-Poisson and response variables 1-8 and 10-12).

Overall, the confidence intervals for the COM-Poisson model were smaller than its counterparts and the point estimates were close among all models. The feral mammal dung covariate had the largest confidence intervals compared to the other covariates. There are still regression coefficients that are very close to zero, suggesting that a better variable selection method may improve the model fit.

\begin{figure}[H]
\vspace{0.35cm}
\setlength{\abovecaptionskip}{.0001pt}
\caption{Correlogram of ant species occurrence from COM-Poisson Model. Stars represent significant correlation at 5\% level}\label{fig:corantcmp}
\includegraphics[width=0.95\textwidth]{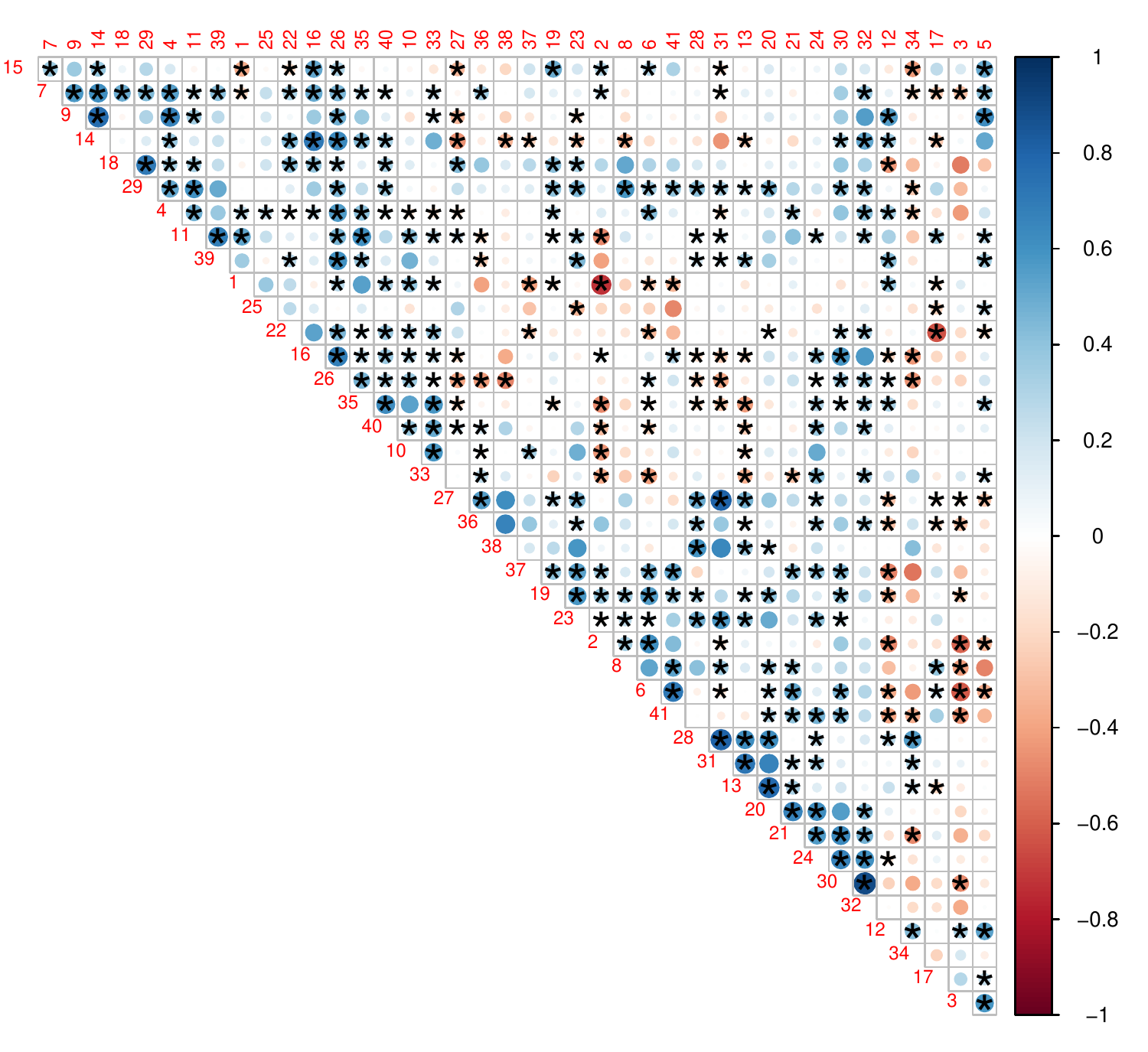}
\centering
\end{figure}

The correlation coefficient for the COM-Poisson model is presented in \autoref{fig:corantcmp}. Among the 820 correlation coefficients returned, COM-Poisson had 333 significant correlation coefficients, NB 278, and Poisson 83. It shows that the correlation coefficient from COM-Poisson had a smaller SE than their counterparts. The stars in the graphic represent the significant coefficients at the 5\% level. The correlation patterns presented in this figure differ somewhat from the ones found in the marginal correlation. It shows the importance of calculating the correlation coefficient in a model that accounts for the effects of the linear predictor. For example, the sample correlation was nearly zero between responses 1 and 2, 2 and 3; while the COM-Poisson correlation between the random effect of these two response variables was negative and significant at 5\% level, for NB only one was significant, and for Poisson neither was significant possibly because of a high SE, once the $\hat\rho$ was equal to -.7 between the random effects of Y1 and Y2, and -.54 between the random effects of Y2 and Y3.

\section{Discussion}

The focus of this article was to propose MGLMM for count data. The great advantage of this model is the ability to deal with multiple outcomes. We specified the framework based on a GLMM with a random intercept, where we structured a joint model whose random effect follows a multivariate normal distribution. As a result, we can measure the variance and correlation of the random effects motivated by the multivariate set of responses. The distributions used for variables of counts were Poisson, NB, and COM-Poisson. The estimation process is the same as a GLMM model. We used the \texttt{TMB} package to implement the model once it provides state-of-art \texttt{C++} libraries that handle automatic differentiation, \texttt{CppAD} \citep{cppad}, linear algebra, \texttt{Eigen C++} \citep{eigen} and parallel computation, \texttt{BLAS} \citep{blas}, among others.

In order to evaluate the properties of the ML estimators, we conducted simulation studies for each distribution, considering three different values for the correlation parameter and four different sample sizes. They were all evaluated through average bias and confidence interval based on the average SE and coverage rate with a nominal level of 95\%. For Poisson distribution, we achieved unbiased and consistent estimators for large samples with intervals for bias ranging at most between (-.2;.2), except for sample size equal to 100 and values for the parameter $\rho$. The coverage rate was close to 95\% in all scenarios considered, varying between 90\% and 98\%. We evidenced a greater variability for NB compared to the Poisson distribution, especially because of the dispersion parameter $\phi$ in NB, necessary to model overdispersion. For NB distribution, we also achieved unbiased and consistent estimators for large samples with bias intervals ranging in most cases between (-.5;.5). In particular, we noticed a greater confidence interval width for the correlation and $\phi_2$ parameter estimates; and small bias for $\sigma_2$ and $\nu_2$ parameter estimates. The coverage rate in most cases was equal to or greater than 95\%; with a coverage rate below 80\% for the correlation parameter when $\rho = 0$ and the sample size equal to 100.

For the COM-Poisson model, the parameter estimators were neither consistent nor unbiased. The regression parameter was underestimated for $\beta_{02}$ and showed no bias for $\beta_{01}$. The correlation parameter $\rho$ was always estimated towards zero: when $\rho=-.5$ it was overestimated, $\rho=+.5$ was underestimated, and for $\rho=0$ it showed no bias. The standard deviation of random effect $\sigma_2$ was overestimated while $\sigma_1$ was underestimated. This behavior may be correlated to the dispersion parameter $\nu$: when more variance was captured from $\sigma_2$ less dispersion was captured from $\nu_2$; on the other hand, when less variance was captured from $\sigma_1$, more dispersion was captured from $\nu_1$.



As most parameters were biased, the coverage rate was not close to 95\% in all scenarios. For $\sigma_2$, $\beta_{02}$ and $\nu_r$ the coverage rate were between 70\% and 90\%; for $\rho$ the coverage rate were close to 80\% in 2 out of 3 scenarios ($\rho=\{-0.5,0.5\}$). The other parameters, $\beta_{01}$, $\sigma_{1}$ and $\rho$ when $\rho=0$ had a coverage rate close to 95\%.


After that, the two datasets were analyzed by each model and variation of them: no correlation, fixed dispersion, fixed variance, and common variance for all random effects. The first dataset was from the AHS survey with five random variables and 5190 participants. According to the fit measures used, the COM-Poisson model provided the best fit. The SE of $\beta$ and $\rho$ estimates were similar among the COM-Poisson, NB, and Poisson models. The $\nu$ parameter captured 4 out of 5 underdispersed response variables and one overdispersed response variable, followed by a small $\sigma$. The COM-Poisson model produced more significant correlation values than its counterparts.

The second dataset was the ANT, which contains 41 response variables that count for the number of ANT species at 30 sites in Australia. The multivariate response can be considered as overdispersed by the GDI. The COM-Poisson model was also the best model regarding \texttt{AIC} and \texttt{logLik}; the model with the best \texttt{BIC} was the COM-Poisson with fixed dispersion. In almost all comparisons, the SEs of the COM-Poisson model was smaller than the other models. The $\nu$ parameter was greater than 1 for all response variables, indicating underdispersion. 

Therefore, we suggest using the MGLMM model framework for count data. In particular, the best results were obtained with the COM-Poisson model in two real datasets. The major advantage of it is the possibility of modeling all response variables at the same time and measuring the correlation between the random effects.

The estimation process of this model was computationally intensive, being a computational challenge to implement the model. For example, the estimation of the COM-Poisson model for AHS data was cumbersome. It took 5 days to fit in a Debian GNU/LINUX 8 (jessie) 92 GB RAM server with an AMD Opteron 6136 processor using two threads. Improving the computational implementation or even using other frameworks are examples of future work. Improving the estimation procedure for the COM-Poisson model in this context is in high demand as well as understanding whether the COM-Poisson distribution can replace the NB distribution for overdispersed data sets.



\bibliography{Silva_AJS}
\end{document}